\documentclass[a4paper,10pt]{article}
 
\usepackage{graphicx}
\usepackage{txfonts}
\usepackage{natbib}
\bibpunct{(}{)}{;}{a}{}{,}

\begin{document}

\begin{centering}
\textbf{\Large Progenitors of type Ia supernovae and the metallicity distribution of G-type dwarfs}\\
\vspace{14pt}
{\large N. Mennekens, D. Vanbeveren, and J.P. De Greve}\\
\vspace{12pt}
\end{centering}
\noindent \emph{\small Astrophysical Institute, Vrije Universiteit Brussel, Pleinlaan 2, 1050 Brussels, Belgium}\\
{\small e-mail: \texttt{nmenneke@vub.ac.be}}
\vspace{10pt}
             
\textbf{Abstract}\\
\emph{Aims.} We investigate the contribution to the formation of type Ia supernovae of the single (a white dwarf accreting from a non-degenerate companion) and double (two merging white dwarfs) degenerate scenario, as well as various aspects of the binary evolution process leading to such a progenitor system. We aim to get a better insight into uncertainties and parameter spaces by means of a combined modelization, resulting in a reduction of the number of possible model approaches. This exclusion of parts of the parameter space is, as will be shown, often independent of uncertainties in the modelization.\\
\emph{Methods.} We use the combination of a population synthesis code with detailed binary evolution and a galactic chemical evolution model to predict the metallicity distribution of G-type dwarfs in the solar neighborhood. Because of the very long lifetime of these stars, this distribution is a good indicator of the entire chemical history of a region. By comparing the observed distribution with those predicted by assuming different type Ia supernova progenitors and evolutionary parameters (e.g. concerning mass and angular momentum loss and common envelope evolution), it is possible to constrain the possible combinations of assumptions.\\
\emph{Results.} We find that in order to reproduce the observed G-dwarf metallicity distribution, it is absolutely necessary to include both the single and double degenerate scenario. The best match is obtained when all merging C-O white dwarfs contribute to the latter. The correspondence is also critically dependent on the assumptions about galaxy and star formation, e.g. the use of the two-infall model vs. a constant star formation rate. However, this does not affect the previous conclusion, which is consistent with the results obtained by investigating type Ia supernova delay time distributions in starburst galaxies.\\

\textbf{Key words:} supernovae: general --
             binaries: close --
             stars: white dwarfs --
             galaxies: starburst --
             Galaxy: abundances --
             Galaxy: evolution --
             Galaxy: solar neighborhood
            
\section{Introduction}

Type Ia supernovae (SNe Ia), the most powerful events produced by nature, are not only of critical importance as cosmological distance indicators (cf. the 2011 Nobel Prize in Physics), but also for the chemical enrichment of galaxies. A large fraction of all iron in the solar neighborhood is known to have been produced by these events. It is commonly agreed upon that SNe Ia originate from white dwarfs (WDs) that exceed a critical mass and as a result undergo a thermonuclear disruption. Many things remain unclear however: whether this happens at the Chandrasekhar mass or above or below, whether only carbon-oxygen (C-O) WDs qualify, whether the disruption is a deflagration or detonation, etc. Perhaps most strikingly, even the type of progenitor system remains uncertain \citep[see][for a review]{maoz2012a}. Since the WD needs to accrete in order to reach the critical mass, it is obvious that only interacting (i.e. binary or multiple) stars can produce such events. However, many different scenarios for the accretion process have been proposed. The most popular ones are known as the single degenerate (SD) and double degenerate (DD) scenario. In the SD scenario \citep[see e.g.][]{whelan1973,nomoto1982}, one WD accretes from a late main sequence (MS) or red giant (RG) companion. The mass transfer rate towards the WD is of critical importance for the scenario to work: if it is too high, the WD will suffer from a RG-like expansion, if it is too low, the WD will burn the aquired mass in classical nova outbursts and will never attain the required critical mass to explode. Hence, a commonly invoked mechanism to explain such a moderate rate is a stabilizing wind from the WD. In the DD scenario \citep[see e.g.][]{iben1984,webbink1984}, the explosion is the result of the merger of two WDs, coming together due to the emission of gravitational wave radiation (GWR). In the case of unequal masses, the least massive of the two will be tidally disrupted and accreted by the more massive one, which is thus allowed to reach the critical mass. It has been previously suggested that such an event would lead to an accretion induced collapse resulting in a neutron star (NS), and not a SN Ia. However, \citet{piersanti2003} showed that the inclusion of the effects caused by rotation may well solve this problem. Recent studies also consider the possibility of subluminous SNe Ia from equal mass C-O WD mergers \citep{pakmor2010}, perhaps even below the Chandrasekhar mass \citep{vankerkwijk2010}. In the DD scenario, one thus needs a double WD binary with a small orbital separation, so that GWR will sufficiently reduce the distance between the stars within the Hubble time. While it is conceivable that due to the nature of most scenarios the explosion mass is always similar, it is very important to address the question of the explosion mechanism, not in the least given the mentioned importance as ``standard candles'' in cosmology.

Many methods have been attempted to discriminate between both scenarios and hence to determine which (or both) of them generates SNe Ia in nature \citep[again, see][for an overview]{maoz2012a}. It then concerns the study of whole populations and of individual explosions, both historic \citep[e.g.][]{schaefer2012} and contemporary \citep[e.g.][the latter two for SN 2011fe]{sternberg2011,dilday2012,chomiuk2012,roepke2012}. A promising method on a global population scale is the study of the delay time distribution (DTD), the response function giving the number of SNe Ia as a function of time after an instantaneous starburst. DTDs can be observationally measured in passively evolving (elliptical) galaxies, and theoretically computed with population synthesis codes, assuming different SN Ia progenitor channels. A comparison of the two then allows to constrain the formation models.

This paper does not address the physical (im)possibility of either formation scenario, but assumes that both of them may work as described by their various authors. From this premise, it is then investigated what scenario (or both) best theoretically reproduces population properties that can be observationally tested, also depending on various other evolutionary assumptions and parameters, both stellar and galactic.

Comparing the observed DTDs of SNe Ia in passively evolving starburst galaxies to those obtained under various assumptions with a population synthesis code, \citet{mennekens2010} (hereafter M10) showed that the SD scenario alone can not account for the amount of SNe Ia required at more than a few Gyr after starburst. Furthermore, it was found that most DD SN Ia progenitors undergo a phase of stable, quasi-conservative Roche lobe overflow (RLOF) followed by a common envelope (CE) phase, as opposed to two successive CE phases. Finally, stellar rotation was proposed to mitigate the discrepancy between the absolute number of events predicted and those observed.

M10 also made a comparison with DTDs previously obtained by other groups using population systhesis methods. The preliminary conclusion was that on many points, a good agreement was found with most of them, however only under the obvious but challenging condition that the assumptions of the other groups were reproduced as much as possible. It was already hinted that, apart from initial distributions, the most important differences in assumptions would be those about mass and angular momentum loss, as well as the treatment of CE phases. Since then, it has become clear to the population synthesis community that it is indeed necessary to investigate the differences in the predictions made by different groups, and to look for the causes thereof. In Sect. 3.4 the macroscopic predictions coming from different implementations of binary evolution into population synthesis codes will be discussed as they are relevant for the predictions concerning single and double WD systems, and hence for potential SN Ia progenitors. A more extended discussion with a comparative analysis is given in \citet{toonen2014}. 

This paper intends to confirm and extend the results of M10 by means of a method independent of the previous, i.e. not only the number of SNe Ia at a certain time (the DTD), but their lasting legacy on the chemical history of a region. In addition, it does not limit itself to the study of passive elliptical galaxies, but aims to reproduce the chemical history and SN Ia rate of the solar neighborhood, in the actively evolving Milky Way Galaxy. There are a number of new uncertainties that come into play when convolving a DTD with a (monolithic) galactic chemical evolution model to reproduce the Galaxy's history. Nevertheless, with this new combined method we are indeed able to confidently exclude a number of combinations of assumptions, and to shed light on which provide promising population-wide properties and thus deserve further scrutiny on a physical level (such as the explosion mechanism). Studies with such methods and for this purpose have indeed been perfomed before, e.g. by \citet{dedonder2004} and \citet{matteucci2009}. These are however subject to improvement, some of them (such as the former) because of significant evolution in the field in recent years, and others (such as the latter) due to the fact that they are not obtained from first principles (see further).

Section 2 summarizes the most important elements contained in the used evolution codes, as well as the assumptions about binary evolution, SN Ia progenitors and the chemical evolution model. Section 3 then discusses the obtained results, while Sect. 4 summarizes the main conclusions.

\section{Assumptions}

\subsection{The Brussels population synthesis code}

The Brussels population synthesis code uses as input thousands of binary evolutionary calculations performed in detail with the \citet{paczynski1967} based Brussels binary evolution code. The latter, under development for over three decades, is extensively described in \citet{vanbeveren1998}. The population code itself is elaborated in \citet{dedonder2004}. We refer to M10 for a specific description of its application to SN Ia progenitors and repeat here only those elements that are of particular importance to this study. One of the most important consequenses of using detailed binary evolution results in the population code is that the effects of accretion on the further evolution of the gainer star are treated in detail. To do this, it is assumed that when a star accretes by direct impact, this process occurs following the ``snowfall'' model by \citet{neo1977}. When, however, the oribtal separation is sufficiently wide to lead to the occurrence of an accretion disk, it has been shown by \citet{packet1981} that the gainer star will relatively soon be spun up to (near-)critical rotation. Therefore, it is assumed that in such case the gainer star will be fully mixed, an event known as ``accretion induced full mixing'' \citep[see][]{vanbeveren1994}. Especially in this case, the further evolution of this star will be markedly different than if it would have been unaffected. Therefore, it is certainly not sufficiently accurate to estimate that the time needed to obtain an accreting WD or a double WD is simply equal to the nuclear lifetime of the least massive star.

The population code starts from a $10^6$ M$_{\odot}$ starburst with a given binary frequency (the fraction of primaries that have a stellar companion) and metallicity Z. Primary star masses $M_1$ are drawn from a \citet{kroupa1993} initial mass function (IMF) normalized between 0.1 and 120 M$_{\odot}$. Three different mass ratio distributions are considered: a flat one (as standard), one favoring high mass ratios \citep{garmany1980} and one favoring small mass ratios \citep{hogeveen1992}, all normalized to allow secondaries with a mass $M_2$ between 0.1 M$_{\odot}$ and $M_1$. Reflecting the results of \citet{abt1983}, the initial orbital period distribution is taken to be logarithmically flat, and is normalized between 1 day and 10 years.

Whenever the initially most massive star in a binary fills its Roche lobe, this will initiate a mass transfer phase. Depending on whether this happens during core H burning (case A), shell H burning (case B) or after core He burning (case C), the nature of this phase and the consequences for the binary system will be different. Many aspects of this mass transfer process are still uncertain and thus characterized by parameters.\\
$\bullet$~In the case of dynamically stable RLOF from donor to gainer, these parameters are the mass transfer efficiency $\beta$ and, if the latter is below unity and thus mass is lost from the system, also the angular momentum loss parameter $\eta$. A commonly made assumption in population synthesis codes is that mass is lost with the specific orbital angular momentum of the gainer star, resulting in small losses of angular momentum and a correspondingly small value of $\eta<<1$. This is however only justified if the matter leaves the system in a way that is symmetric with respect to the equatorial plane of the gainer (e.g. through an enhanced stellar wind or bipolar jets). As there is limited evidence for significant mass loss in such a way in intermediate mass non-degenerate stars, our standard model is that matter will instead form a non-corotating circumbinary ring after passing through the second Lagrangian point, resulting in a much larger angular momentum loss and $\eta=2.3$ \citep[see][]{soberman1997}.\\
$\bullet$~If the donor's outer layers are already deeply convective by the time mass transfer starts, this will result in an unstable CE phase. Formalisms to describe this have been proposed by \citet{webbink1984} and \citet{nelemans2000}, both parameterized by an energy conversion efficiency, termed $\alpha$ and $\gamma$ respectively.

For details on how either mass transfer process is treated in the Brussels code (e.g. the possible range of the parameters), as well as how the two are distinguished, we refer to M10. At a later point in the evolution, after the originally most massive star has become a WD, the originally least massive star will also fill its Roche lobe, initiating a mass transfer episode in the other direction. Because of the fact that the accreting star is a WD with a small surface, and because the mass ratio between the two stars is often very large, it is assumed that this will always lead to an unstable episode, hence it is modeled as a CE phase. The only exceptions are the SD SN Ia progenitors discussed in the next subsection.

\subsection{Type Ia supernova progenitors}

In this study, mainly the two most popular formation channels for SNe Ia are considered, the single degenerate (SD) and double degenerate (DD) scenario. Various others, including the so called core degenerate channel \citep{kashi2011}, various delayed detonation models \citep[e.g.][]{hachisu2012}, etc. have tentatively been included in our population synthesis code, but none of them has proven to be able to result in a notable number of SNe Ia at the right time (see Sect. 3). Some other scenarios which have been considered as promising, such as the sub-Chandrasekhar `double detonation' SD model involving He-rich donors \citep[see e.g.][considered this scenario in population synthesis context]{livne1990,ruiter2011} and the C-O WD `violent mergers' DD model \citep[see e.g.][]{ruiter2013}, are not yet considered here in great detail. Preliminary results indicate that they are in our simulations unable to greatly change the picture. A more detailed study of this, as well as the influence of core growth, He-star evolution and WD accretion efficiencies \citep[for an illustration of their importance, see][]{bours2013} on our present results will be reserved for a future paper. Interestingly, the progenitor parameter spaces of SD and DD events do not overlap, implying that it is possible that both formation channels are at work simultaneously. The population code normally assumes that both scenarios do indeed work in union, however it also allows to ``turn off'' either one, so that the implications of only one (or altered) scenario can be explored.

\subsubsection{Single degenerate scenario}

For the SD scenario progenitors, we do not explicitly calculate the mass accretion on the WD ourselves, but we make use of the results of the groups that treated this scenario in detail. When in a binary star consisting of a non-degenerate star and a WD, the former fills its Roche lobe, \citet{hachisu1999,hachisu2008} identified those zones (dependent on the WD mass) in the (companion mass, orbital period)-plane for which the mass transfer rate will fall within the narrow zone in which stable accretion up to the critical mass (assumed to be the Chandrasekhar mass of 1.4 M$_{\odot}$) is possible. In addition to these results for Z=0.02, \citet{kobayashi1998} also calculated the zones for Z=0.004, so that combination allows to interpolate for any Z.

\subsubsection{Double degenerate scenario}

Concerning the DD scenario, it is assumed that every merger of two WDs, together meeting or exceeding the Chandrasekhar limit of 1.4 M$_{\odot}$, will result in a SN Ia at the time that GWR has reduced the distance between both to zero. Because the exact physics of the explosion are as yet unclear and it is widely believed that the scenario only works for the merger of two C-O WDs, our computations are also performed with this additional restriction. Pursuant to recent speculation on the matter, we also investigate the influence on the population assuming sub-Chandrasekhar C-O WD mergers to result in SNe Ia.

It is of importance to repeat that there are two typical ways in which a double WD, eventually resulting in a DD SN Ia, can be formed. To obtain a double WD in a binary, the primary and secondary need to fill their Roche lobes successively, resulting in two mass transfer phases. The first of these two phases can either be a stable RLOF event (either conservative or non-conservative) or result in an unstable CE evolution. The second phase is assumed to result always in a CE phase, since the accretor is a WD. Detailed descriptions of either evolution channel as well as typical examples are given in M10. The conclusion is that double WDs having gone through two successive CE phases produce DD SNe Ia already after a few hundred Myr, while events having undergone a RLOF event have a delay time of up to several Gyr, but require the RLOF to be (quasi-)conservative to avoid merging of the components.

\subsection{The Brussels galactic evolution model}

The galactic code combines different output sets of the population code with a galaxy formation model to self-consistently calculate the star formation rate (SFR) and the subsequent chemical evolution of a galaxy. To provide sufficient room for interpolation, the chemical enrichment (including SNe Ia) by a starburst is included on the one hand for a population of single stars, on the other for a population of 100\% binaries, and this for four different metallicities: Z=0 (obtained by extrapolation), 0.0002, 0.002 and 0.02. It concerns a full set of yields as discussed by \citet{dedonder2004}, based on those of \citet{woosley1995} but in particular, as motivated in the former paper, with the iron core-collapse yields divided by a factor of two. Combination with the current SFR, binary fraction and metallicity thus allows to calculate the current chemical abundances and SN Ia rate at any time. As was already found by \citet{dedonder2004}, one of the most important ways in which binaries significantly alter the chemical history of a galaxy compared to the computations with single stars only, is by the chemical enrichment (especially in iron) caused by SNe Ia. Therefore, it is critical that the rate of these events be an integral part of every chemical evolution model, which is physically sound and consistent with other ingredients, and not just ``imported'' from observations, let alone a free parameter.

Concerning the binary frequency, there are two possible assumptions: either this value is assumed to be constant in space and time, or it is treated as a function of metallicity. The latter assumption, a binary frequency that linearly increases with Z, does not produce any models that satisfactorily reproduce the chemical evolution of the solar neighborhood (as will be shown in Sect. 3). Therefore, a constant binary frequency is assumed, with a value of 70\%. This reasonably high value is required in order to produce sufficient SNe Ia (which can obviously only occur in binaries) to attain the observed SN Ia rate and the necessary SN Ia-specific chemical (iron-)enrichment. This is consistent with the results of M10, where it was found that a very high binary frequency is also required (and not even sufficient) to explain the absolute number of SNe Ia observed in elliptical (starburst) galaxies.

Another important ingredient of any galactic evolution code is a galaxy formation model. This describes not only how the galaxy itself was formed, but more critically also provides a way to calculate the SFR as a function of time. The evolution of the SFR critically determines the chemical composition, SN Ia rate, etc. An often used galaxy formation model is the two-infall model by \citet{chiappini1997}. This model assumes that the Milky Way Galaxy was formed by two successive gas infall phases, the first forming the thick disk (including halo and bulge), the second the thin disk. The gas infall rate in either phase is given by the following relation:

\begin{equation}
\frac{d\sigma_{g,inf}(t)}{dt}=Ae^{-t/\tau_1}+Be^{-(t-t_{max})/\tau_2}.
\label{eq:gasinfall}
\end{equation}

In principle, this equation obviously also depends on the distance to the galactic center, however since we only consider the solar neighborhood we are only concerned with applying it there. $\tau_1$ and $\tau_2$ are the mass accretion timescales for the halo and disk phase, which are taken as 2 Gyr and 7 Gyr (at the position of the Sun) respectively. $t_{max}$ is the time of maximum accretion on the thin disk, and is equal to 1 Gyr. The values of the scaling parameters $A$ and $B$ are determined by the requirement of the galaxy formation model to reproduce the current observational constraints. The SFR is assumed to depend on both the gas surface density $\sigma_g(t)$ and the total mass surface density $\sigma(t)$. \citet{chiosi1980} adapted the prescription of \citet{talbot1975} to the infall model:

\begin{equation}
\Psi(t)=\nu\left(\frac{\sigma(t_{now})}{\sigma(t)}\right)^{k-1}\left(\frac{\sigma_g(t)}{\sigma(t_{now})}\right)^k
\label{eq:SFR}
\end{equation}

\noindent where $\nu$ is the star formation efficiency in Gyr$^{-1}$. Both this parameter and $k$ can be different during the first and second infall phase. $\Psi(t)$ is set to 0 if $\sigma_g(t)$ is below 7 M$_{\odot}$/pc$^2$. The favored model takes values of respectively 2 and 1 Gyr$^{-1}$ for the efficiencies, and $k=1.5$ in both phases. It should be noted that the values of the parameters described here and used in this paper are taken from \citet{matteucci2009} and differ from those in the original two-infall model by \citet{chiappini1997} which were used in previous papers by the Brussels group. Notably, the `new' values result in a SFR that is much less concentrated at early times, and instead is more flattened out.

A second possible galaxy formation model that is considered in this paper is simply a constant SFR. The onset of star formation is assumed to be delayed by 0.5 Gyr, which is the required time to allow the surface gas density to rise to 7 M$_{\odot}$/pc$^2$. After that, the SFR is constant with its value determined by the currently observed gas and star densities. This turns out to be about 4 M$_{\odot}$/pc$^2$Gyr, resulting in a total star formation roughly equal to the case of the two-infall model.

In both cases, the infalling primordial gas has a composition of X=0.76, Y=0.24 and Z=0.

Since a similar study by \citet{dedonder2004}, studies of the same kind have been performed by \citet{greggio2008}, \citet{matteucci2009} (albeit both not with internally computed SN Ia DTDs, but with various adopted ones) and \citet{kobayashi2011} (albeit for the SD scenario only). The former three concluded that the best match between prediction and observation was found when both the SD and DD model were combined.

\subsection{Observational galactic parameters}

A critical prediction made by galactic evolution models is the abundance of various chemical elements. These are usually expressed logarithmically, relative to hydrogen (H), and normalized to the solar abundance. In this context, the most important abundance is that of the element iron (Fe). For the chemical yields of a single SN Ia event, those of the W7 model by \citet{iwamoto1999} are assumed, both for SD and DD progenitors. Most importantly, this includes a yield of 0.626 M$_{\odot}$ Fe per SN Ia.

The age-metallicity relation (AMR) is the evolution of [Fe/H] (= log(Fe/H) - log (Fe/H)$_{\odot}$) as a function of time. Stars that are formed at a certain time are obviously assumed to contain the metallicity of that instant. Therefore, if these stars are still observable today, and their age can be determined, they provide a way to reconstruct the chemical history of the Galaxy. Stars that are very suited for this purpose are the G-type dwarfs (0.80M$_{\odot} \le M \le 1.05$M$_{\odot}$), since they have an extremely long lifetime of up to the age of the Galaxy itself. Therefore, the metallicity distribution of these local G-type dwarfs, i.e. the relative number of observed G-dwarfs with respective [Fe/H], is indicative of the entire chemical history of the solar neighborhood. Since the DTDs of SD and DD SN Ia progenitors are markedly different, this chemical history should be critically influenced by which scenario is at work (or both).

The observational G-dwarf metallicity distributions that will be used for comparison in this paper are those by \citet{holmberg2007}. Two distributions are derived in this paper, one for a spherical solar neighborhood and one for a cilindrical one. Since in our galactic code the solar neighborhood is defined as a cilindric region of 1 kpc around the Sun, mainly the latter will be used. It should be noted that a more recent analysis of the Geneva-Copenhagen survey data by \citet{casagrande2011} results in a slightly different G-dwarf metallicity distribution, which has the same morphological shape but peaks at a slightly higher value of [Fe/H]. As the difference ($\sim 0.1$ dex) is not too large, and the \citet{holmberg2007} distribution provides the advantage of having been explicitly calculated also for a cilindrical solar neighborhood, the latter will be used for comparison. Additionally, if theoretical models are ruled out by being unable to produce enough Fe, these conclusions will not be affected if an even higher [Fe/H] must be matched.

While the G-dwarf metallicity distribution seems quite robust (those already mentioned are quite similar to the ones by \citet{lee2011}, and also numerous others found in the literature over the past decade), the same is not true for the AMR. As will be elaborated on in Sect. 3.2, the recent literature contains a wide variety of such observed relations, depending on the method of their construction. Some of these AMRs show a rising trend in [Fe/H] as a function of time, while others do not. Also the width of the AMR (i.e. the extent of scatter in [Fe/H] at a given age) and its possible origins is a matter of debate. For this reason, our comparison between prediction and observation will focus mainly on the G-dwarf metallicity distributions themselves, and less on the AMR.

\subsubsection{Comparison method}

The current age of the Galaxy is taken to be 13.2 Gyr, the time after which the theoretically obtained star and gas surface densities are compared to the currently observed ones. The latter are taken from \citet{calura2010} and references therein, yielding 37.5 $\pm$ 10 M$_{\odot}$pc$^{-2}$ for the stellar surface density and 10.5 $\pm$ 3.5 M$_{\odot}$pc$^{-2}$ for the gas surface density. Other parameter-sensitive output that can be compared to current values are a SFR of 3.5 $\pm$ 1.5 M$_{\odot}$pc$^{-2}$Gyr$^{-1}$ \citep[according to the same source's analysis of][]{rana1991} and a local SN Ia rate of 0.003 $\pm$ 0.002 pc$^{-2}$Gyr$^{-1}$ \citep[according to the analysis of][]{cappellaro1996}. The latter corresponds to 3 $\pm$ 2 SNe Ia per millennium in the Milky Way Galaxy. Only when these values are reproduced to within the observational uncertainty is a model considered as possibly valid. The final distribution that then needs to be checked, the actual goal of the study, is the G-dwarf metallicity distribution.

Using the same \citet{kroupa1993} IMF as the population code, the galactic code calculates from the SFR how many G-dwarfs are being born at any given time. These are obviously assumed to contain the metallicity that is prevalent at the time of their birth. Taking into account the mass-dependent lifetime of each G-dwarf, it is then calculated which fraction of those born at each timestep is still observable today. The [Fe/H]-distribution of these remaining stars is then plotted in a histogram with bin size 0.1 dex, ready to be compared to observations. Theoretically, every star born at the same time thus has the same Fe-content. As mentioned before, however, observations indicate that this may not be the case in reality \citep{ramirez2007,holmberg2007,casagrande2011}. It may thus prove to be necessary to supplement the theoretically obtained [Fe/H] at each moment with a random (positive or negative) deviation, e.g. distributed following a certain standard deviation. A possible reason for this ``intrinsic scatter'' in the AMR is the radial migration of stars over large galactocentric distances \citep{roskar2008}. The possible implications of this open question for the results of this study will be discussed in Sect. 3.2.

It is thus clear that the true AMR and G-dwarf distribution are subject to a significant number of (possible) influences which are not taken into account by simple, monolithic, galactic chemical evolution models such as our own (or, e.g., the one used by \citet{matteucci2009}). This does not mean, however, that the latter cannot be used to add to our understanding or to constrain evolutionary assumptions. This research aims not at validating a certain model. On the contrary, we want to exclude those which are clearly incompatible with observations. And that within a degree that most likely exceeds the effect of any possible unaccounted uncertainty. The influence of radial migration and related effects, as studied by e.g. \citet{roskar2008} and \citet{schoenrich2009}, may be severe for the morphological shape and width of the G-dwarf metallicity distribution. However, for all models in these studies the shift in location of the peak of this distribution is not larger than 0.1 dex.

\section{Results and discussion}

Anticipating the results in this section, we find that reproducing the observed DTD is a necessary but not sufficient condition for a certain combination of assumptions to satisfactorily reproduce the observed G-dwarf metallicity distribution. Therefore, the first test will again focus on those DTDs. Only when the DTD is in agreement with the observed ones can there be hope of reproducing the observed G-dwarf metallicity distribtution in a second step. A DTD corresponding to observations is however no guarantee for success, not only because of galactic evolutionary considerations, but also because of the limited timeframe during which the theoretical DTD can be compared with observation: at early times no observational test for the DTD is available, while this part obviously also contributes to the galactic chemical evolution. It should be remarked that, while convenient, there are no a priori reasons why a model that is able to produce a DTD corresponding to those observed in elliptical galaxies, should simultaneously also be able to reproduce on a galactic level the chemical history of our own solar neighborhood. These two observational tests are entirely independent. The fact that this will indeed turn out to be the case (for the first time from first-principle assumptions) is reassuring concerning the robustness of the model and the level of (unaccounted for) uncertainties.

Additionally, whether a model with satisfactory DTD will also reproduce the observed G-dwarf metallicity distribution critically depends on the Z-sensitivity of the model. The DD channel is relatively independent of metallicity, its efficiency even increasing a bit with lower Z \citep[similar conclusions in this regard were recently obtained by][]{meng2012}. However, the number of events produced by the SD channel is decimated when Z is lowered from 0.02 to 0.002, and reduces to zero for even lower Z. This is of course a result of the WD wind no longer being able to stabilize the mass stream. This means that a model which solely relies on the SD channel to produce SNe Ia with short delays, will only be able to provide these events (and their Fe enrichment) after the metallicity has already been enhanced. Hence, the early evolution of the Galaxy will be characterized by Fe-deficiency.

\subsection{Delay time and G-dwarf metallicity distribution}

In this subsection we will first establish our standard model of parameters and other assumptions, as also used in M10. It will be shown that this model, with neither progenitor scenario nor with both combined, is satisfactory, even with the suggestions for improvement made in that paper. Thus, we will investigate what changes in the assumptions would be required to mitigate the discrepancy between prediction and observation. Subsequently, the following will be studied:

\begin{itemize}
\item the influence of angular momentum assumptions ($\alpha$ vs. $\gamma$-scenario for CE, values of parameters, modeling of angular momentum loss)
\item whether inclusion of any non-traditional SN Ia progenitor scenarios can significantly change the picture
\item the influence of uncertainties intrinsic to the tradiational progenitor scenarios (what is the best value for $c_1$ in SD systems, are DD SNe Ia restricted to C-O WD mergers, do sub-Chandrasekhar C-O WD mergers also explode?)
\item all along, also the influence of galactic evolutionary parameters and assumptions (e.g. galaxy formation model, binary frequency type and value) will be studied.
\end{itemize}

An overview of all considered combinations of scenarios, assumptions and parameters, as well as their ability to reproduce the observed DTD and G-dwarf metallicity distribution, is given in Table \ref{tab:result}.

\begin{figure*}
\centering
   \includegraphics[width=12cm]{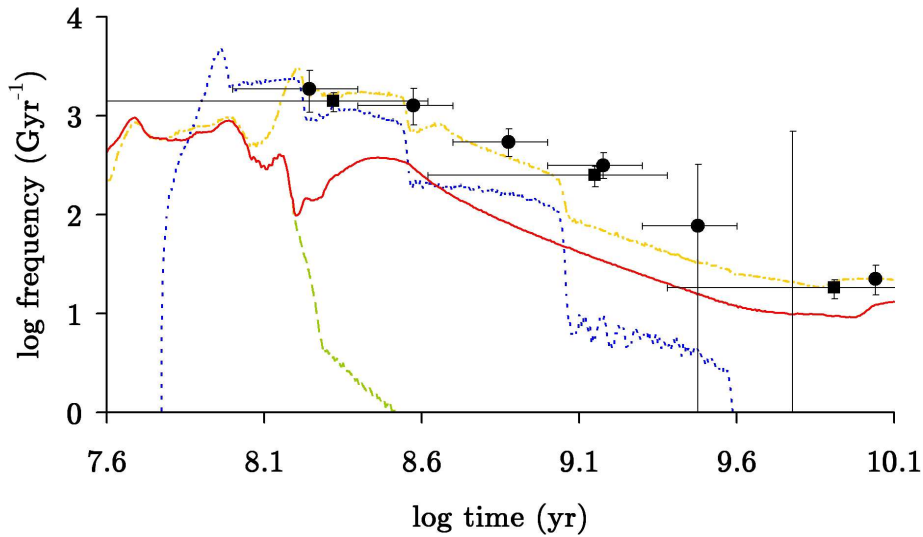}
     \caption{DTDs obtained with the DD scenario for $\beta_{max}=1$ (solid) and $\beta_{max}=0$ (dashed), as well as with the SD scenario for $\beta_{max}=1$ and $c_1=3$ (dotted). Combined SD + DD model DTD with $\beta_{max}=1$ and $c_1=1$, but with a convective core mass increase of 10\% (dashed-dotted). Observational data points by \citet{totani2008} (circles) and \citet{maoz2012b} (squares).}
     \label{fig:DTDold}
\end{figure*}

\begin{figure*}
\centering
   \includegraphics[width=12cm]{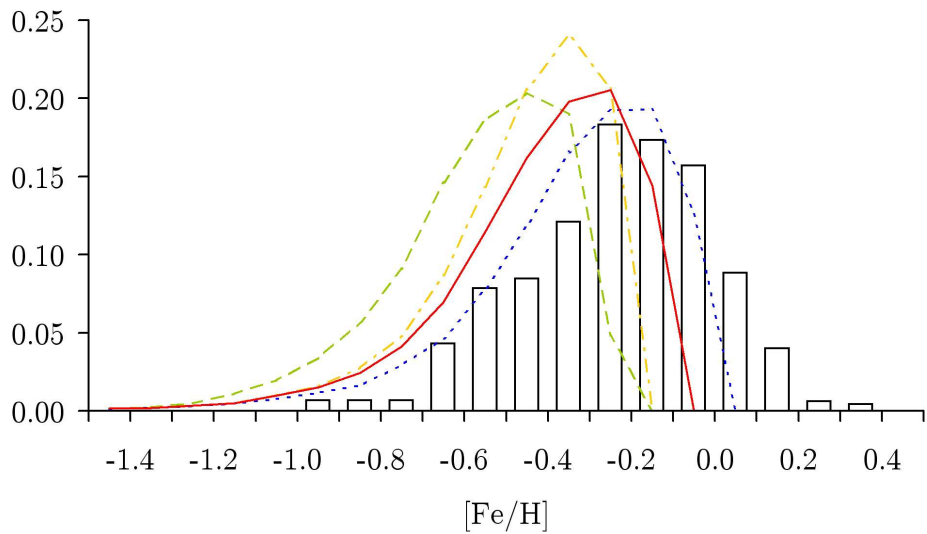}
     \caption{G-dwarf metallicity distributions obtained with the combined SD (with $c_1=3$) + DD model either without (solid) or with (dotted) 10\% convective core mass increase, as well as for the SD (with $c_1=3$, dashed) respectively DD (dashed-dotted) scenario alone. Observational data for a cilindrical solar neighborhood (white histogram) by \citet{holmberg2007}.}
     \label{fig:Gold}
\end{figure*}

\begin{figure*}
\centering
   \includegraphics[width=12cm]{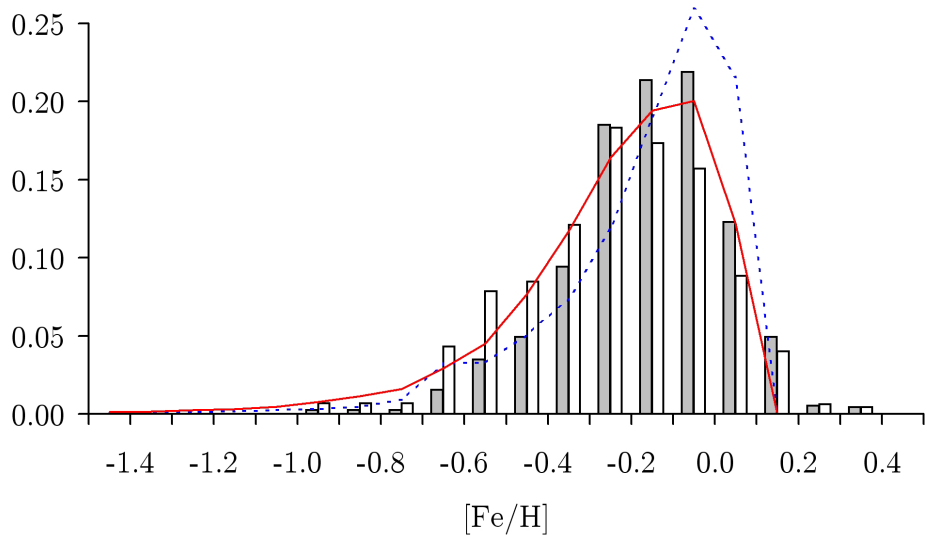}
     \caption{G-dwarf metallicity distributions obtained with the combined SD (with $c_1=1$) + DD model, with SN Ia rates multiplied by a factor of 2.5, for a constant SFR (solid) and for the two-infall model (dotted). Observational data for a spherical (gray histogram) and cilindrical (white histogram) solar neighborhood by \citet{holmberg2007}.}
     \label{fig:Gnew}
\end{figure*}

\begin{figure*}
\centering
   \includegraphics[width=12cm]{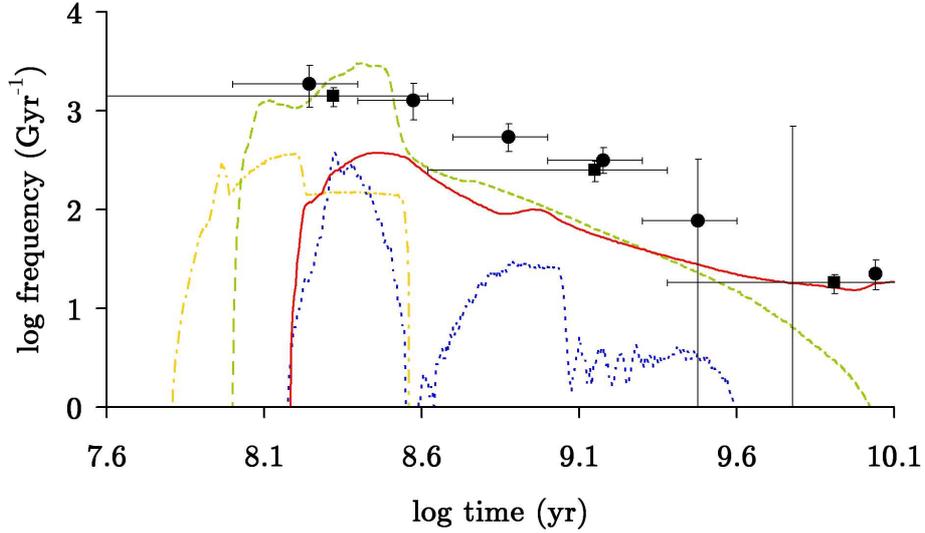}
     \caption{DTDs obtained with the $\gamma$-scenario for CE-evolution through the DD (solid) and SD (dotted) channel. DTDs obtained with the $\alpha$-scenario, using the parameter values determined by \citet{dewi2000}, through the DD (dashed) and SD (dashed-dotted) channel. Observational data points by \citet{totani2008} (circles) and \citet{maoz2012b} (squares).}
     \label{fig:DTDnl}
\end{figure*}

\begin{figure*}
\centering
   \includegraphics[width=12cm]{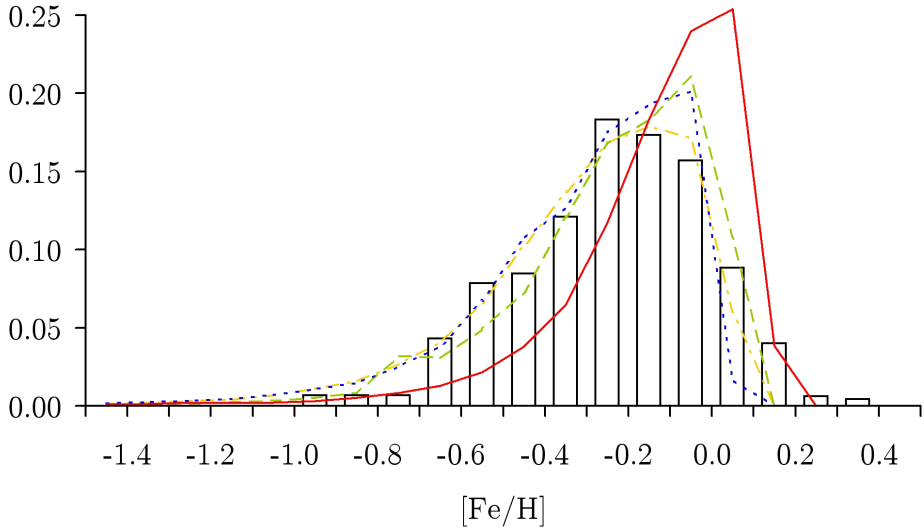}
     \caption{G-dwarf metallicity distributions obtained with the combined SD (with $c_1=3$) + DD model, all with SN Ia rates multiplied by a factor of 2.5, for the $\alpha$-formalism with parameter values by \citet{dewi2000} (solid), for the $\gamma$-formalism and either a constant SFR (dotted) or the two-infall model (dashed), as well as for the standard $\alpha$-formalism but with only super-Chandra CO-CO WD mergers resulting in DD SNe Ia (dashed-dotted). Observational data for a cilindrical solar neighborhood (white histogram) by \citet{holmberg2007}.}
     \label{fig:Gnl}
\end{figure*}

\begin{figure*}
\centering
   \includegraphics[width=12cm]{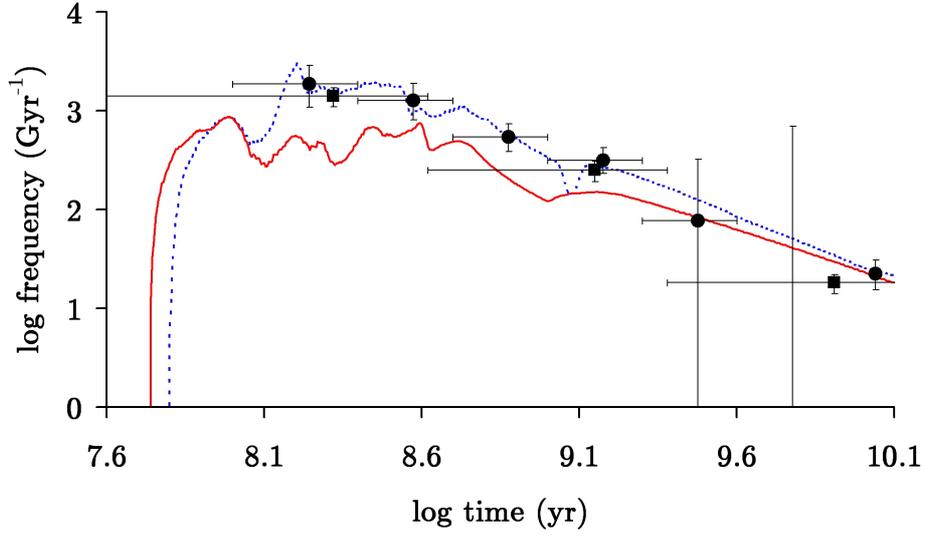}
     \caption{DTD obtained with the assumption that all double C-O WD mergers result in SNe Ia, through the DD (solid) channel. Combined SD (with $c_1=1$) + DD model DTD under the same assumption, but with a convective core mass increase of 10\% (dotted). Observational data points by \citet{totani2008} (circles) and \citet{maoz2012b} (squares).}
     \label{fig:DTDwd}
\end{figure*}

\begin{figure*}
\centering
   \includegraphics[width=12cm]{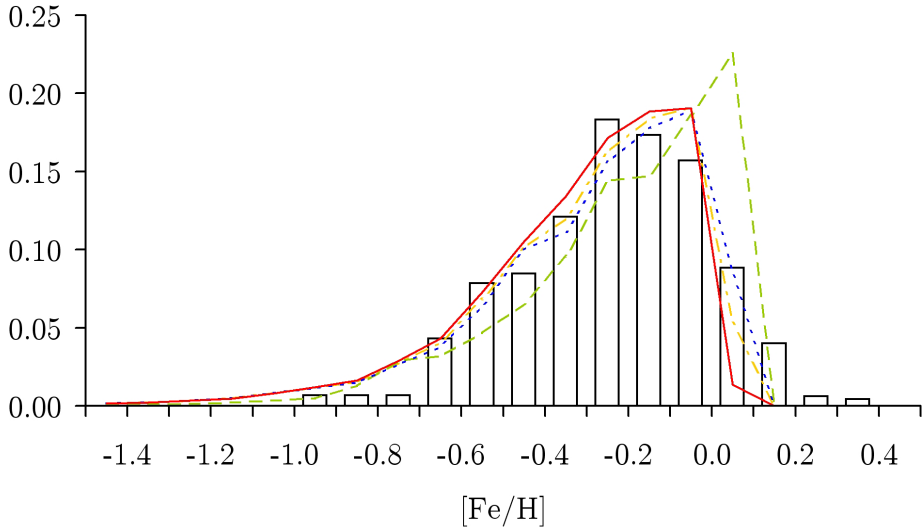}
     \caption{G-dwarf metallicity distributions obtained with the combined SD + DD model, with $c_1=3$ under the assumption that all double C-O WD mergers result in SNe Ia (solid), with $c_1=1$ and the same assumption but for a convective core mass increase of 10\% for either a constant SFR (dotted) or the two-infall model (dashed), as well as for a constant SFR but with SN Ia yields scaled to their total explosion mass if $<$ 1.4 M$_{\odot}$ (dashed-dotted). Observational data for a cilindrical solar neighborhood (white histogram) by \citet{holmberg2007}.}
     \label{fig:Gwd}
\end{figure*}

\begin{figure*}
\centering
   \includegraphics[width=12cm]{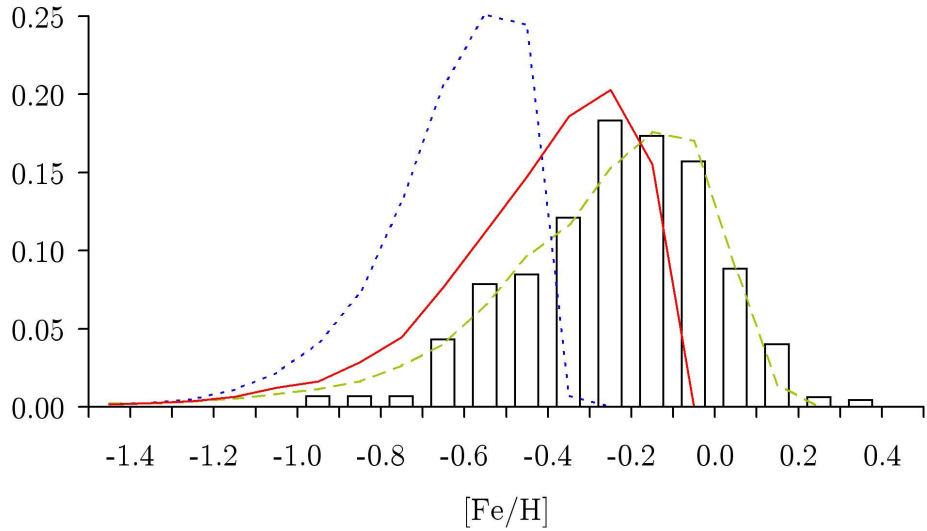}
     \caption{G-dwarf metallicity distributions obtained with the combined SD (with $c_1=1$) + DD model, with all double C-O WD mergers resulting in SNe Ia and a convective core mass increase of 10\%, for a constant binary frequency of 35\% (solid), for a binary freqency increasing linearly with Z (dotted) and for the standard binary fraction of 70\% but with added artificial scatter as explained in the text (dashed). Observational data for a cilindrical solar neighborhood (white histogram) by \citet{holmberg2007}.}
     \label{fig:fb}
\end{figure*}

\begin{figure*}
\centering
   \includegraphics[width=12cm]{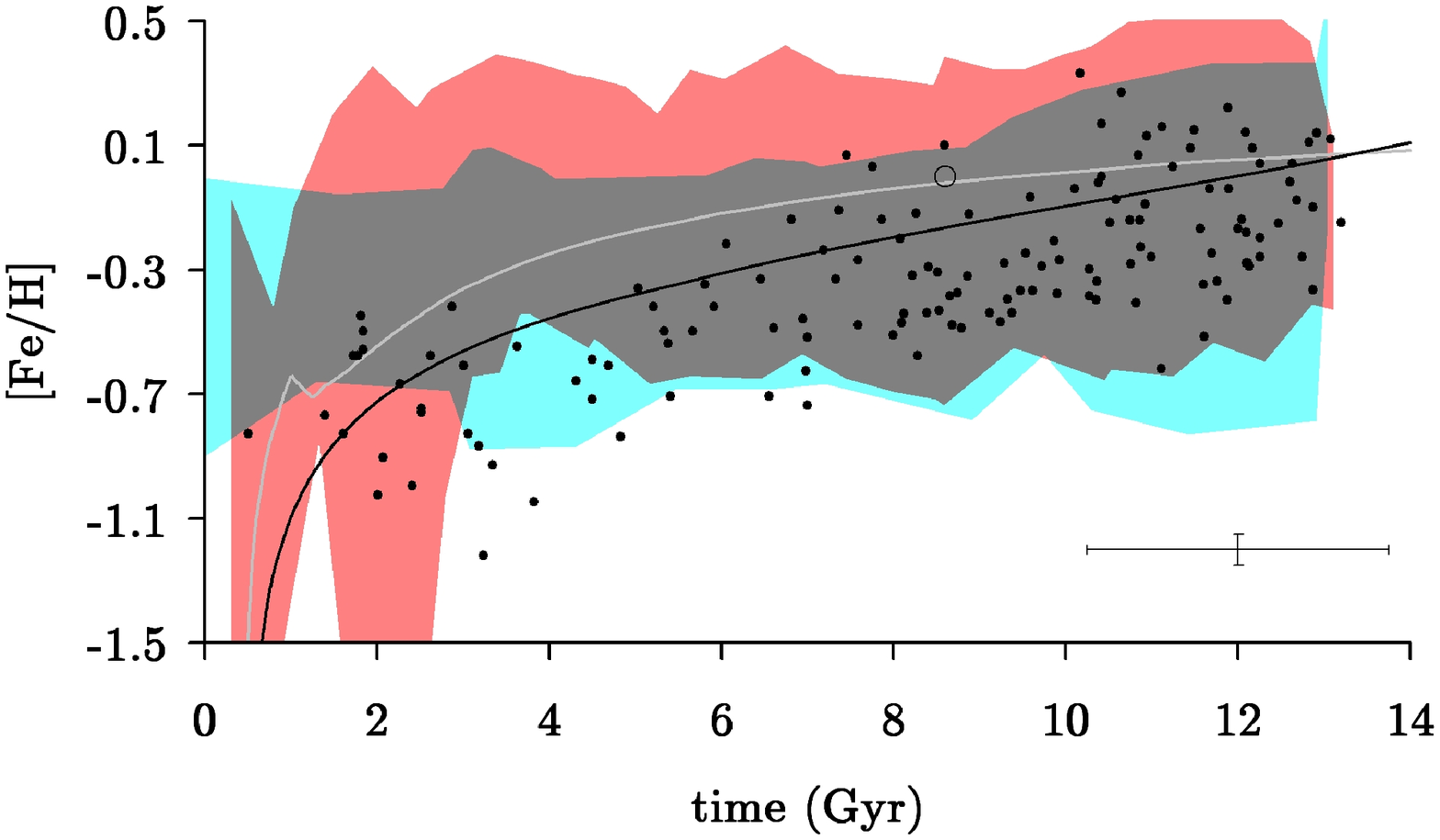}
     \caption{AMR with the combined SD (with $c_1=1$) + DD model, with all double C-O WD mergers resulting in SNe Ia and a convective core mass increase of 10\%, for a constant SFR (black) and for the two-infall model (gray). Observational data points by \citet{ramirez2007} (dots) and for the Sun (open cricle), as well as zones (shaded) in which the data points by \citet{holmberg2007} (lower) and \citet{casagrande2011} (upper) lie, but see text for important note on their representativeness. At the lower right, the typical observational error bar for center of the plot (increasing towards the left) is shown.}
     \label{fig:AMR}
\end{figure*}

\begin{figure*}
\centering
   \includegraphics[width=12cm]{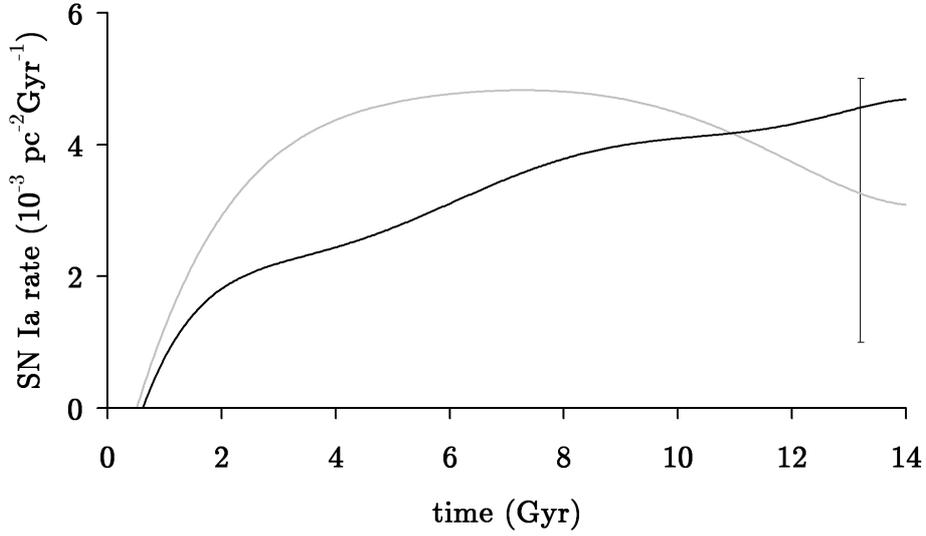}
     \caption{SN Ia rate with the combined SD (with $c_1=1$) + DD model, with all double C-O WD mergers resulting in SNe Ia and a convective core mass increase of 10\%, for a constant SFR (black) and for the two-infall model (gray). Currently observed rate is indicated by error bar.}
     \label{fig:IaRate}
\end{figure*}

\begin{figure*}
\centering
   \includegraphics[width=12cm]{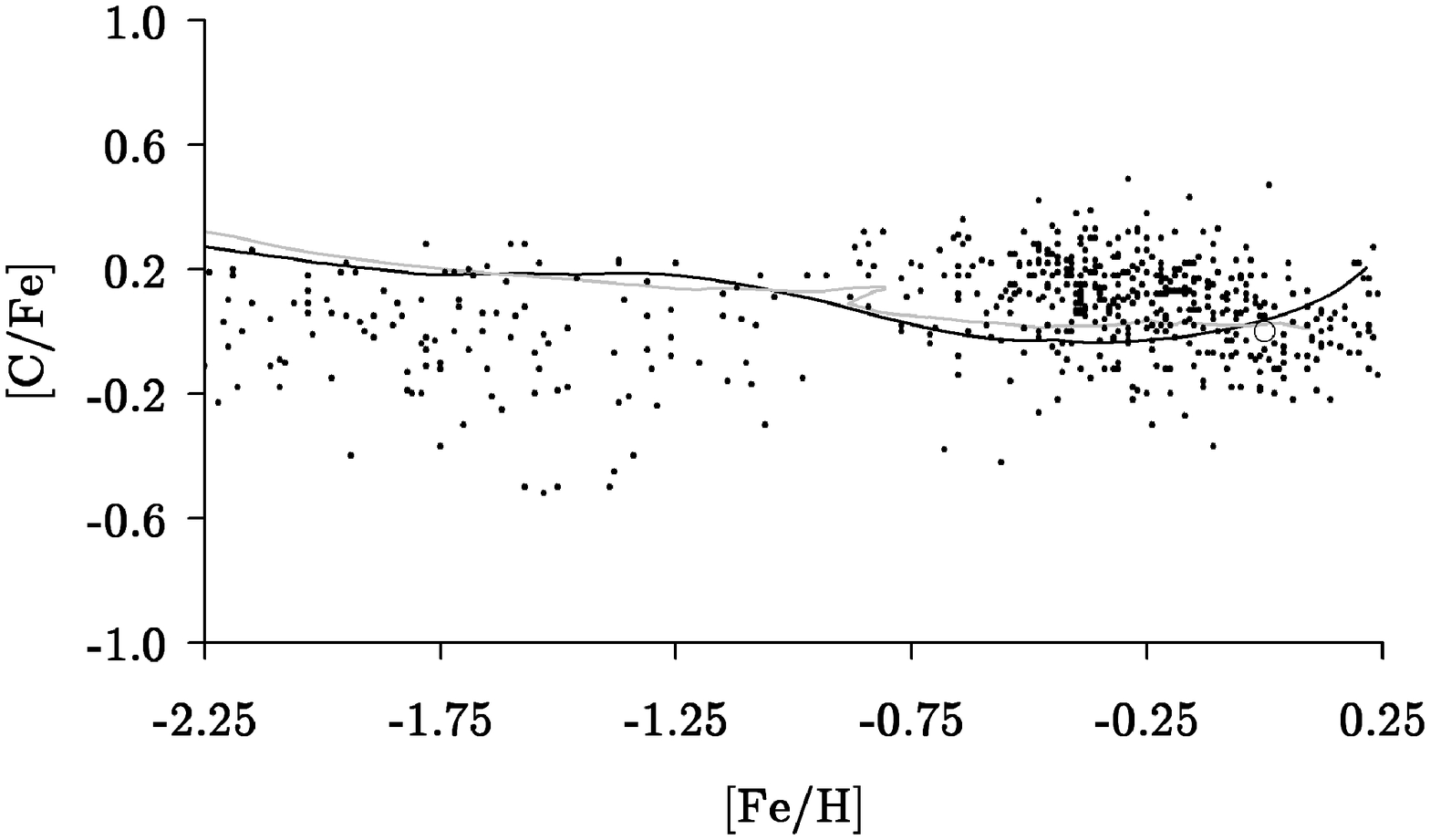}
     \caption{[C/Fe] vs. [Fe/H] with the combined SD (with $c_1=1$) + DD model, with all double C-O WD mergers resulting in SNe Ia and a convective core mass increase of 10\%, for a constant SFR (black) and for the two-infall model (gray). Observational data by \citet{chiappini2003}, \citet{reddy2003}, \citet{bensby2006} and \citet{fabbian2009} (dots) and for the Sun (open cricle).}
     \label{fig:C}
\end{figure*}

\begin{figure*}
\centering
   \includegraphics[width=12cm]{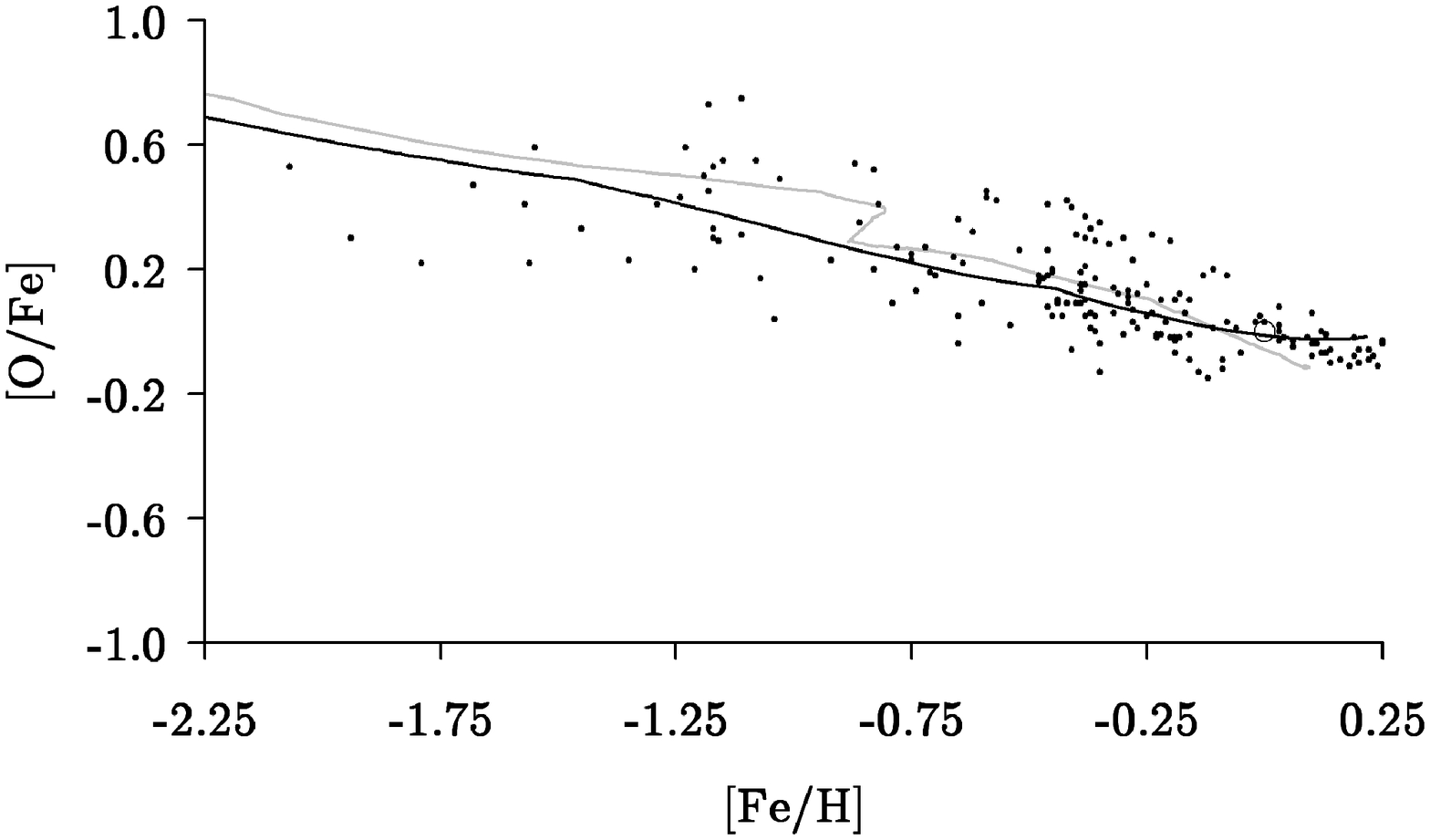}
     \caption{[O/Fe] vs. [Fe/H] with the combined SD (with $c_1=1$) + DD model, with all double C-O WD mergers resulting in SNe Ia and a convective core mass increase of 10\%, for a constant SFR (black) and for the two-infall model (gray). Observational data by \citet{gratton2003}, \citet{reddy2003} and \citet{calura2010} (dots) and for the Sun (open cricle).}
     \label{fig:O}
\end{figure*}

\begin{figure*}
\centering
   \includegraphics[width=12cm]{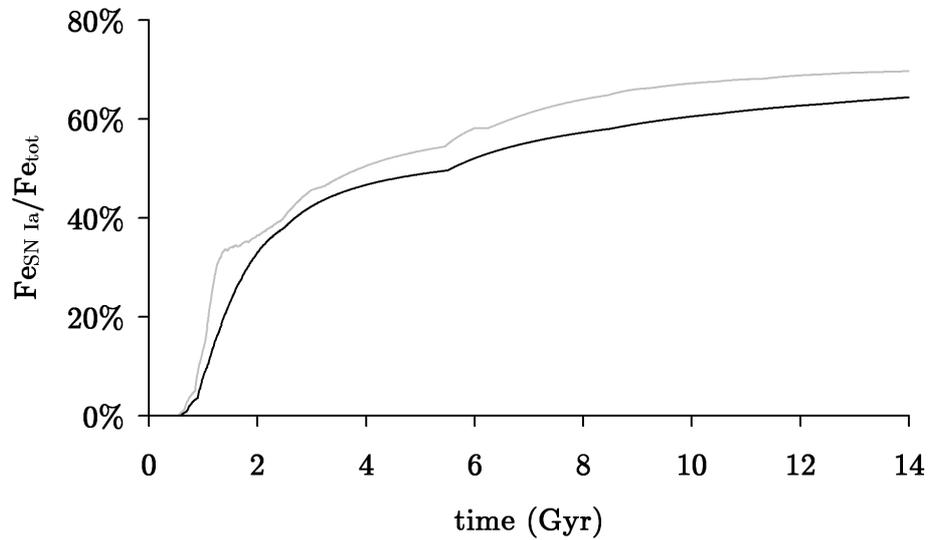}
     \caption{Fraction of all Fe produced by SNe Ia with the combined SD (with $c_1=1$) + DD model, with all double C-O WD mergers resulting in SNe Ia and a convective core mass increase of 10\%, for a constant SFR (black) and for the two-infall model (gray).}
     \label{fig:Fe}
\end{figure*}

\begin{table*}
\centering
\tiny
\begin{tabular}{c c c c c c c c}
\hline
SN Ia model(s) & mass \& AM & CE & other & DTD & DTD & G-dwarf & G-dwarf \\
 & loss & evol. & & match? & Fig. & match? & Fig. \\
\hline
SD($c_1=3$) & $\beta=1$, $L_2$ & $\alpha\lambda=1$ & & no & \ref{fig:DTDold} & no & \ref{fig:Gold} \\
DD & $\beta=1$, $L_2$ & $\alpha\lambda=1$ & & shape & \ref{fig:DTDold} & no & \ref{fig:Gold} \\
DD & $\beta=0$, $L_2$ & $\alpha\lambda=1$ & & no & \ref{fig:DTDold} & no & \\
SD($c_1=3$) & $\beta=0$, $L_2$ & $\alpha\lambda=1$ & & no & & no & \\
SD($c_1=1$) + DD & $\beta=1$, $L_2$ & $\alpha\lambda=1$ & $M_{CC}$+10\% & yes & \ref{fig:DTDold} & no & \ref{fig:Gold} \\
SD($c_1=3,10$) + DD & $\beta=1$, $L_2$ & $\alpha\lambda=1$ & $M_{CC}$+10\% & yes & & no & \\
SD($c_1=0-10$) + DD & $\beta=1$, $L_2$ & $\alpha\lambda=1$ & & shape & & no & \ref{fig:Gold} \\
SD($c_1=1,3$) + DD & $\beta=1$, $L_2$ & $\alpha\lambda=1$ & x2.5 & yes & & yes (cSFR) & \ref{fig:Gnew} \\
\hline
SD($c_1=1,3$) + DD & $\beta=1$, $L_2$ & $\gamma\alpha$ & & no & \ref{fig:DTDnl} & no & \\
SD($c_1=1,3$) + DD & $\beta=1$, $L_2$ & $\gamma\alpha$ & x2.5 & yes & & yes & \ref{fig:Gnl} \\
SD($c_1=1,3$) + DD & $\beta=1$, $L_2$ & Dewi et al. & & no & \ref{fig:DTDnl} & no & \\
SD($c_1=1,3$) + DD & $\beta=1$, $L_2$ & Dewi et al. & x2.5 & yes & & yes (cSFR) & \ref{fig:Gnl} \\
SD($c_1=1,3$) + DD & $\beta=1$, O & $\alpha\lambda=1$ & & shape & & no & \\
SD($c_1=1,3$) + DD & $\beta=0$, O & $\alpha\lambda=1$ & & shape & & no & \\
\hline
SD($c_1=1,3$) + DD(only CO $>M_{Ch}$) & $\beta=1$, $L_2$ & $\alpha\lambda=1$ & & shape & & no & \\
SD($c_1=1,3$) + DD(only CO $>M_{Ch}$) & $\beta=1$, $L_2$ & $\alpha\lambda=1$ & x2.5 & yes & & yes (cSFR) & \ref{fig:Gnl} \\
CD & any & any & & no & & no & \\
delayed detonation SD & any & any & & no & & no & \\
He-star channel & any & any & & no & & no & \\
SD($c_1=1,3$) + DD(all CO) & $\beta=1$, $L_2$ & $\alpha\lambda=1$ & & yes & \ref{fig:DTDwd} & almost & \ref{fig:Gwd} \\
\emph{SD(c$_1=$ 1$,3$) + DD(all CO)} & \emph{$\beta=$ 1, L$_2$} & \emph{$\alpha\lambda=$ 1} & \emph{$M_{CC}$+10\%} & \emph{yes} & \emph{\ref{fig:DTDwd}} & \emph{yes (cSFR)} & \emph{\ref{fig:Gwd}} \\
previous, with DD scaled if $<M_{Ch}$ & $\beta=1$, $L_2$ & $\alpha\lambda=1$ & $M_{CC}$+10\% & yes & \ref{fig:DTDwd} & yes (cSFR) & \ref{fig:Gwd} \\
SD($c_1=1,3$) + DD(all $>M_{Ch}$ + all CO) & $\beta=1$, $L_2$ & $\alpha\lambda=1$ & & yes & & yes (cSFR) & \\
any previous with SD or DD off & any & any & & no & & no & \\
any previous & any & any & $f_b<0.7$ or Z-dep. & no & & no & \ref{fig:fb} \\
\hline
\end{tabular}
\caption{Different models discussed in Sect. 3.1 and their correspondence to observation, as well as corresponding figures (best model, to which Figs. 9-13 apply, in italics). Parts delimited by horizontal lines denote different subsections. See text for meaning of symbols and abbreviations.}
\label{tab:result}
\end{table*}

\subsubsection{The canonical SD and DD scenario}

The DTDs obtained with the population code for a single starburst consisting of 100\% binaries (as is the case for all DTDs in this paper) are shown in Fig. \ref{fig:DTDold}. Represented are the SD DTD for $\beta_{max}=1$ and $c_1=3$, as well as the DD DTDs for $\beta_{max}=1$ and 0. Superimposed are the (unscaled) observational DTDs of \citet{totani2008} \citep[including one data point by][]{mannucci2005} and \citet{maoz2012b}, converted into the same units as explained in M10. The SD DTD for $\beta_{max}=0$ barely differs from the one shown. Other assumptions are Z=0.02, $\alpha\lambda=1$ and a flat mass ratio distribution. As already concluded by M10, independent of the latter choices, the SD DTD drops away too fast and too soon to keep matching the observations after a few Gyr, even with values for $c_1$ up to 10. For $\beta_{max}=1$ the DD distribution matches the observations in morphological shape, but lies a factor 2-3 too low in absolute value at the 11 Gyr point. The drastic decrease in the DD DTD for $\beta_{max}=0$ means two things: firstly, the events which disappear in the latter case (about 80\% of all DD SNe Ia) have all gone through a stable RLOF event and thus not two CEs; secondly, in order to be able to reproduce the morphological shape of the observations, $\beta_{max}$ needs to be close to 1, i.e. $\geq 0.9$. Also shown in the figure is the combination of both SD and DD, albeit with $c_1=1$ and assuming a 10\% increase in convective core mass due to rotation (as proposed by M10), the latter mainly enhancing the DD channel. This combination matches the observed DTDs both in shape and number, but as will be seen later this does not automatically mean that inserting this DTD in the galactic model will yield a satisfactory galactic chemical evolution outcome.

Fig. \ref{fig:Gold} shows the G-dwarf metallicity distribution obtained with only the SD and DD channel respectively. The SD rates have been calculated with $c_1=3$, a moderately high value which results in rates only marginally smaller than if the maximum value of $c_1=10$ would be chosen. While the DD DTD is the only one still compatible with the observations in morphological shape, it is obvious that the G-dwarf metallicity distribution does not correspond to observations. The same is true to an even more dramatic extent for the SD channel. In none of the performed simulations, with none of the stellar or galactic evolutionary parameter choices, we obtain a situation in which either scenario alone succeeds in even marginally reproducing the observed G-dwarf metallicity distribution, even when taking into account the uncertainties on the results introduced by the oversimplifications of a monolithic galactic evolution model.

As does Fig. \ref{fig:DTDold}, M10 showed that when a 10\% increase in convective core mass is assumed, the combined SD + DD model matches the observational DTDs well. However, it is obvious from Fig. \ref{fig:Gold} that this is not true for the corresponding G-dwarf metallicity distribution. This shows that the distribution peaks at too low [Fe/H] values, indicating that not enough Fe is formed early on in the galactic evolution. This conclusion stands independent of the chosen value for $c_1$. The reason is that the early part of the DTD, prior to the first observational test, does not provide enough Fe pollution soon after the starburst. These early events are caused through the SD channel (which is not much affected by the convective core mass increase and does not work for low Z anyway), as well as through WD mergers already massive enough without the increase. As will be shown later, if early events would also increase by a factor of 2-3, the G-dwarf metallicity distribution would match observations. Obviously, the distribution for the combined SD + DD model but without convective core mass increase (also shown in Fig. 2) results in an even poorer match with observations.

As an academic exercise, Fig. \ref{fig:Gnew} shows the G-dwarf metallicity distribution which is obtained when the standard model SD and DD channels (from Fig. \ref{fig:DTDold}) are combined, but multiplied by a factor of 2.5 over the entire time range\footnote{To allow for straightforward comparison, the same factor of 2.5 will be used in several subsequent models that otherwise underestimate the observed Fe-abundances.}. In that case, the DTD perfectly matches the observations. When the standard two-infall model is used, this results in a G-dwarf metallicity distribution that peaks in the correct location, but with much less spread than the observations. On the other hand, when a constant SFR is assumed, the predicted G-dwarf metallicity distribution shows a good match with the observed one. As a side note, it is remarkable that the one obtained with the two-infall model shows a better resemblance with the observational distribution obtained for a spherical solar neighborhood as is used for comparison by \citet{kobayashi2011}.

\subsubsection{Mass and angular momentum loss considerations}

The DTDs and the resulting G-dwarf metallicity distribution were also calculated using the $\gamma$-scenario for CE evolution in the first mass transfer phase (i.e. the one during which the companion is not yet a compact object), and the $\alpha$-scenario in the second. The results are shown in Fig. \ref{fig:DTDnl} and \ref{fig:Gnl} respectively. The same figures also show the results obtained with the $\alpha$-scenario for CE evolution, but with the $\alpha\lambda$ parameter not set to the same fixed value for all stars but instead calcultated separately for each system according to the results of \citet{dewi2000}, resulting in values for this parameter much smaller than one, on average $\alpha\lambda\approx0.25$. It can be surmized that for both alternative CE treatments the conclusions are similar as for the standard $\alpha$-scenario: in order to reproduce the observed G-dwarf metallicity distribution, the absolute number of SN Ia events needs to be multiplied by 2.5. This then also results in a DTD matching the observed one. Only when using the $\alpha\lambda$-values by \citet{dewi2000} does this multiplication result in a slight Fe-overabundance, indicating that a lower multiplication factor is needed there for the best match. Also notable is that when using the $\gamma$-scenario, the morphological shape of the G-dwarf metallicity distribution is acceptable for both the two-infall model and the constant SFR.

Use of the specific gainer orbital angular momentum loss assumption instead of the $L_2$ one (see Sect. 2.1) results for $\beta_{max}=1$ in SD and DD SN Ia rates changing by not more than some 10\%, nor is the morphological shape of the DTD severely altered. The same is thus also true for the resulting G-dwarf metallicity distributions and the conclusions from those. One possibly important difference is that with this angular momentum loss assumption, in the case of $\beta_{max}=0$, the distributions are not greatly influenced compared to $\beta_{max}=1$, owing to the very limited angular momentum loss. Thus, much lower values of $\beta$ than with the $L_2$ assumption still yield DTDs compatible in morphological shape with the observations.

\subsubsection{Non-standard progenitor models}

When the standard SD + DD model is considered, but with DD SNe Ia restricted to the merger of two C-O WDs (and still exceeding the Chandrasekhar limit), the total number of SNe Ia goes down by some 20\%, but this does not compromise the morphological shape of the DTD. The corresponding G-dwarf metallicity distribution, again for the DTD multiplied by a factor of 2.5, is also shown in Fig. \ref{fig:Gnl}. It demonstrates that to obtain a satisfactory distribtution, the multiplication factor then merely has to be increased slightly.

One of the non-standard SN Ia formation scenarios that has been tested is the core-degenerate (CD) channel \citep[][]{kashi2011}. For this, it is assumed that the merger of a WD with the core of a non-degenerate star, together exceeding the Chandrasekhar mass, will result in a SN Ia. This merger can happen at any time between the formation of the first and second WD, but most likely during a CE phase. According to \citet{ilkov2012}, the SN Ia does not necessarily take place at the time of merger: if they have a rapid rate of rigid rotation, the explosion of merger products with a mass below 1.48 M$_{\odot}$ can be delayed by a significant time (up to 10 Gyr), until magnetic braking has provided enough downspin. This can not only deliver the events with long delay time, but can also explain the absence of H in the spectrum of the explosion. It should also be noted that the CD channel is relatively independent of the $\alpha\lambda$ parameter of CE evolution. If this product is set to 0.1 instead of 1, the CD rate changes by only -15\%, compared to -95\% and +80\% for SD and DD respectively. Also the corresponding change in morphological shape of the distribution is much smaller than for those two. However, as was already shown by M10 (without making reference to SNe Ia) the number of WD + non-degenerate mergers is extremely large immediately after starburst, but then drops very fast (with the number of such events sharply falling by several orders of magnitude after 0.3 Gyr). Although the total number of these potential CD events matches the total number of SNe Ia quite well \citep[as noted by][]{ilkov2013}, we thus find that the number of early events ($<$ 0.3 Gyr) lies much too high to match the observed DTD, while the number of late events ($>$ 0.4 Gyr) lies much too low. The mentioned delay between merger and explosion does not solve this problem, as it can only apply to systems with a total mass $<$ 1.48 M$_{\odot}$, which represent only 10\% of the total. Also the G-dwarf metallicity distribution obtained under the CD assumption is unable to match the observed one: owing to the extremely high number of events (up to $>10^{-2}$ Gyr$^{-1}$M$_{\odot}^{-1}$, which does make it the only model in this entire study able to match the extremely high data point from the Magellanic Clouds study by \citet{maoz2010}) at early times, it results in a strong Fe-overproduction regardless of other assumptions. These conclusions also stand under the additional restriction that the WD involved must be of the C-O type.

For the same reason, the delayed detonation SD model proposed by \citet{yoon2004}, \citet{distefano2011} and \citet{justham2011}, for which a comprehensive progenitor parameter space was calculated by \citet{hachisu2012}, is unable to provide a significant increase in SNe Ia with long delay in our simulations. This scenario assumes that rotating WDs can accrete matter until highly exceeding the Chandrasekhar limit, not exploding until the star has spun down. However, only for the small minority of rigidly rotating WDs with a mass below 1.5 M$_{\odot}$ can this cause a significant delay.

Recently there has been much speculation about the possibility of sub-Chandrasek\-har C-O WD mergers exploding as SNe Ia \citep[see e.g.][]{badenes2012}. If it is so that all sub-Chandrasekhar C-O WD mergers explode as SNe Ia, the number of DD events increases by a factor of four compared to only including super-Chandra C-O WD mergers. This is of course subject to a physical model of how such an event can explain the observed properties, with such efforts underway by e.g. \citet{vankerkwijk2010} and \citet{pakmor2010}, mainly for nearly equal-mass WDs. In our simulations, the average total mass of sub-Chandra C-O WD mergers is 1.24 M$_{\odot}$ (not too far below the Chandrasekhar mass) and the average mass ratio 0.70 (not too far below unity and as high as in super-Chandra mergers). Fig. \ref{fig:DTDwd} shows the DD DTD obtained under the assumption that all C-O WD mergers, regardless of their total mass, indeed result in such an event. This DTD (in combination with the SD channel from Fig. \ref{fig:DTDold}) shows a very good agreement with the observational ones. However, as Fig. \ref{fig:Gwd} shows (still with the W7 yields even for sub-Chandra events), the obtained G-dwarf metallicity distribution shows not enough Fe enrichment, even though the discrepancy is not too large. The reason for this is again that at the very start of the DTD, before any comparison with observation is available, the SN Ia rate is low, especially for Z$<$0.02 (DTDs not shown). \citet{wang2009} propose a He-star channel which they claim leads to very early SNe Ia through the SD channel. When introduced into our population code, this channel produces SNe Ia between 44 and 130 Myr after starburst, however only amounting to 1.8\% of the total number of events in the standard SD and DD scenario. One hypothetical solution to the situation is when one does not restrict the DD SN Ia explosions to C-O WDs. If, in addition to assuming that all C-O WD mergers explode (regardless of total mass), one also assumes that all super-Chandrasekhar mergers explode (regardless of composition), then the early-time events (starting as early as 20 Myr) are provided by DD mergers where the more massive star is an ONeMg WD (this explains the difference in early behaviour between the DD DTDs from Figs. \ref{fig:DTDold} and \ref{fig:DTDwd}). Without compromising the shape and number of the DTD at times for which observations are available, the obtained G-dwarf metallicity distribution then shows an excellent match with the observations.

A much less artificial model, that also reproduces the observed G-dwarf metallicity distribution, is when the assumption about all C-O WD mergers exploding as SNe Ia is combined with the assumption about the 10\% convective core mass increase. The then obtained DTD is also shown in Fig. \ref{fig:DTDwd} and perfectly matches the observations, without multiplication factor. The same is true for the corresponding G-dwarf metallicity distribution shown in Fig. \ref{fig:Gwd}. The `dip' in the DTD of this best model around 1 Gyr can be explained as follows: the sudden decrease at this time is a result of the SD channel suddenly becoming much less efficient (as can be seen from Fig. \ref{fig:DTDold}). The increase a short time later is due to the DD channel suddenly becoming more efficient (as is also the case for the DD DTD without core mass increase shown in Fig. \ref{fig:DTDwd}). This is because, under the assumption that all C-O WD mergers explode, at this time there is a sudden turn-on of systems with an initial secondary mass between 1.2 and 2.6 M$_{\odot}$, which are very numerous. It should also be noted that while all DTDs in this paper are shown for a starburst binary frequency of 100\%, the results would not be greatly affected if 70\% were taken instead (as for the solar neighborhood galactic model used to calculate the G-dwarf metallicity distributions). Specifically, this `best model' DTD of Fig. \ref{fig:DTDwd} would then still traverse all the observational error boxes. Figure \ref{fig:Gwd} shows in fact three different distributions obtained under these assumptions. Firstly for the flat SFR vs. the two-infall model, both for $c_1=1$. In models with flat SFR, this is the only $c_1$-value for which stellar surface density, gas surface density and current SN Ia rate are within observational constraints at the same time. The model with $c_1=3$ results in a slightly too high SN Ia rate (the same was true in the case of the combined SD + DD model multiplied by a factor of 2.5). In calculations assuming the two-infall model, a greater range of $c_1$ meets these constraints but, as noted, these have more trouble reproducing the morphological shape of the G-dwarf metallicity distribution. The third distribution shown in Fig. \ref{fig:Gwd} is also for a flat SFR, but with all DD SN Ia yields scaled to their total explosion mass (if below 1.4 M$_{\odot}$), i.e. taking into account the fact that a sub-Chandrasekhar DD SN Ia may well produce less Fe (and other yields) than a Chandrasekhar mass one. The figure shows however that this open question does not critically affect the metallicity distribution. Figure \ref{fig:Gwd} shows a slight Fe-overproduction when using the SFR from the two-infall model. In that case, the assumption that all C-O WD mergers result in DD SNe Ia yields a better G-dwarf metallicity distribution without the additional assumption of a 10\% convective core mass increase than with it. However, it then only concerns the location of the peak of the distribution, its morphological shape is still reproduced better with the flat SFR. Finally, it should also be remarked that even with the 10\% convective core mass increase and the assumption that all C-O WD mergers result in SNe Ia, neither the DTD nor the G-dwarf metallicity distribution obtained through the DD channel alone reproduce the observational distributions.

As an illustration of the previously claimed requirement to have a high and constant binary frequency, Fig. \ref{fig:fb} shows the same best model but for a binary frequency of 35\% (half the standard value), respectively one varying linearly with Z. It is obvious that the G-dwarf metallicity distribution in both cases shows a very strong underproduction of Fe.

Even in the case of the best model, the predicted G-dwarf metallicity distribution shows too little (i.e. no) stars with very high metallicity, and too many with very low metallicity. This is the case for all reasonably simple theoretical galactic chemical evolution models, and is long known as the G-dwarf problem. A possible way to mitigate the situation is by assuming that stars are not necessarily observed at the location of their birth (and for which their [Fe/H] is thus representative). Such a migration scenario \citep{roskar2008} allows to explain the observational presence of stars with higher metallicity than can be produced locally according to the model. Additionally, it allows to reproduce to a certain extent the intrinsic scatter in the AMR that is observed (i.e. the fact that not all stars born at the same time have the same [Fe/H] in reality, see also next subsection). A simple simulation to mimic the effects of the intrinsic spread on the results of our monolithic model is shown in Fig. \ref{fig:fb}. It has been obtained from the result shown for the best model and constant SFR (and should thus be compared with the dotted line in Fig. \ref{fig:Gwd}) by assuming that only two-thirds of the G-dwarfs assigned to a certain bin of width 0.1 dex are actually observed in that bin, while one-sixth are actually observed in the next bin, and one-sixth in the previous. This result shows that, as expected and also shown by \citet{roskar2008} in the context of the migration-induced scatter, it slightly lowers the peak of the distribution and increases its width. In our case, the location of the peak is obviously not affected, whereas in the results of \citet{roskar2008} it shifts by not more than 0.1 dex towards higher metallicity (as higher Z material migrates from areas closer to the Galactic center). However, the influence of this effect on the location of the peak (and thus the mean value) of the metallicity distribution is clearly much smaller than the typical influence of changing from one assumption to the other in our above considerations. Assuming that there are no other effects of even greater importance which have not yet been taken into account in galactic evolution studies, the main conclusions of this study, obtained by eliminiating those models which significantly under- or overestimate the Fe-enrichment of the Galaxy, will thus not be affected. The smaller trends noticed in our results may be overwhelmed by effects such as this and are thus not necessarily significant. Our comparisons between the two-infall model and a constant star formation rate (two opposite extremes) also show that the galaxy formation model can be important for the fine-tuning of the model, but that the two agree quite well on which combinations of assumptions are certainly excluded and which are promising.

\subsection{Other galactic evolutionary constraints}

Figure \ref{fig:AMR} shows the AMR for the model which best approximates the observed G-dwarf metallicity distribution, i.e. the SD (with $c_1=1$) + DD model but with all C-O WD mergers exploding as SN Ia and a convective core mass increase of 10\%. This is done both for a constant SFR and for the SFR following from the two-infall model. It should be noted that distributions for other (reasonably) satisfactory models, such as the same model without the core mass increase or the standard SD + DD model with SN Ia rates (artificially) enhanced by a factor of 2.5, look very similar. When the predicted AMR is compared to the observational ones obtained by \citet{holmberg2007} and \citet{casagrande2011}, this is not very instructive. The latter namely suggest an AMR that is almost flat, but has a very significant and intrinsic scatter. As a result of the latter, all theoretical models produced in this study lie within the observationally populated zone, but none of them is able to reproduce an AMR with this general (lack of) shape. This is because the infalling pristine gas is unable to fully compensate for the large Fe-enrichment caused by SNe Ia. If the \citet{holmberg2007} or \citet{casagrande2011} AMR is indeed representative for the solar neighborhood as well as intrinsic (i.e. not caused by contamination effects due to dynamics), and thus the one to be reproduced, this presents a serious problem for theoretical predictions. As already mentioned in the context of the G-dwarf distribution, almost no theoretical chemical evolution models are able to produce an AMR that is flat throughout most of the Galaxy's history, although again the radial migration assumption may help in this respect. However, an alternative AMR was derived by \citet{ramirez2007} under different conditions on the star selection and especially the age uncertainty, as well as different stellar parameter zero points (especially temperature, leading to different ages). This observational AMR does indeed show a rise in function of time, as is predicted by theoretical evolution models. However, it is likely that this sample consisting of stars suitable for high-resolution high S/N spectral analysis is not representative for the solar neighborhood, as it is based on a collection of spectroscopic data for a very limited number of stars, selected in a non-random fashion, but poorly constrained nonetheless (I. Ram\'irez, personal communication, 2012). Nevertheless, this AMR is also shown in Fig. \ref{fig:AMR}. Whether or not the \citet{holmberg2007} or \citet{casagrande2011} AMR are indeed representative for the solar neighborhood additionally depends on the validity of the used stellar isochrones. For completeness, it should also be noted that if one were to construct a G-dwarf metallicity distribution from the stars selected by \citet{ramirez2007} for the construction of their AMR, this G-dwarf metallicity distribution would be extremely more spread out than the one by \citet{holmberg2007} or \citet{casagrande2011}. While such distribution would be questionable due to the mentioned non-representative nature of the dataset, it is of interest to note that to retrieve such a spreaded G-dwarf metallicity distribution in our predictions, the intrinsic scatter added to the time-dependent [Fe/H] must be taken very large, to the extent where it would correspond to the entire scatter observed in the AMR.

The evolution of the SN Ia rate as a function of time is shown in Fig. \ref{fig:IaRate}, with the currently observed rate indicated. It shows an acceptable level of agreement with observation for either model, although it should be remembered from Fig. \ref{fig:Gwd} that only for the flat SFR a good agreement in morphological shape was found for the G-dwarf metallicity distribution.

To check that they are not incompatible with observation, Figs. \ref{fig:C} and \ref{fig:O} show the C and O abundance variation obtained with the same model, again for both flat SFR and the two-infall model. While early [C/Fe]-values seem to lie slightly too high, and the decreasing trend between [Fe/H]=-0.5 and 0 suggested by observations is not reproduced, there is no significant discrepancy. The early evolution ([Fe/H]$<$-1.5) of both elements can be greatly modified by making different assumptions about direct black hole formation and hypernova yields \citep[as described in][]{dedonder2003}, but this does not affect the G-dwarf metallicity distribution which is the object of this study.

As a final step, the fraction of Fe in the solar neighborhood originating from SNe Ia is investigated. Fig. \ref{fig:Fe} shows its evolution as a function of time, for the constant SFR, respectively two-infall model calculations. Between 65 and 70\% of the Fe present in the solar neighborhood today was created by SNe Ia.

\subsection{Comparison to previous results}

\citet{dedonder2004} were the first to conclude that the predicted G-dwarf metallicity distribution best matches the observed one when both SD and DD events contribute. However, this was before the time that a reliable observational DTD was available, and did thus not allow a comparison in that respect. Moreover, the extent of the parameter studies was more limited, and galactic observational constraints and SN Ia progenitor assumptions have strongly evolved over the last decade. In the meantime, \citet{matteucci2009} came to the same conclusion of SD+DD ``probably being the most realistic one''. However, this was achieved on the one hand with DTDs not calculated self-consistently from first-principle assumptions but instead imported either from observations or very simple models, and on the other hand with an artifical multiplication factor ensuring a match between theory and observation (a bit like our illustrative academic exercice in Fig. \ref{fig:Gnew}). This is particularly troublesome given the earlier mentioned critical sensitivity of DTDs to the metallicity. If one, Z-independent DTD is imported into the chemical evolution model and used throughout, this chemical evolution model will obviously not account for the (very important) dependence of SN Ia rates on metallicity.

One significant difference between the current study and the results of \citet{dedonder2004} on a galactic level is that we now find that a much flatter SFR is required than in the original two-infall model by \citet{chiappini1997}. While this is one of the results in this study to which the caveat applies that the influence on the G-dwarf metallicity distribution is perhaps not significantly greater than that of possible uncertainties, it is in line with the more recent results of the Trieste group, who altered the parameters of the two-infall model to indeed flatten out the SFR. The main conclusion from 2004 remains: both SD and DD scenarios are required to contribute. However, as explained in detail in Sect. 3.1, the ease with which the observational G-dwarf metallicity distribution is then reproduced has now greatly been reduced.

\subsection{Other considerations}

As mentioned in the Introduction, concurrent with this study a comparison has been made by \citet{toonen2014} between different binary population synthesis models. Apart from the results published there, this exercise has made us aware of a number of new ingredients in the Brussels code, other than parameterizations we were aware of, which may have a profound influence on the outcome of the evolution of a certain system. We will now investigate their consequences for the results obtained so far. A first point is the wind mass loss during the AGB phase. Its yields as computed by \citet{vandenhoek1997} are included in the code. However, we do not follow the angular momentum loss as a result of these winds in detail. Instead, the mass decrease from terminal age main sequence to post-mass transfer phase (i.e. the C-O core) and the resulting orbital separation variation is calculated in one step, thus in fact acting as if this mass is lost in the CE. Of course, this results in a different orbital separation for individual systems as in those codes which follow the AGB phase in detail throughout at the expense of more computation time. However, the effect of this simplification on the eventual DTDs (and thus also the G-dwarf metallicity distributions) has been found to be negligible. The same is true for the simplification of not following in detail the convective core mass evolution between systems filling their Roche lobe early and late on the AGB. The importance thereof will however be further explored in more detail in a future paper.

\section{Conclusions}

We have tested the ability of the two most popular type Ia supernova progenitor scenarios, the single and double degenerate channel, to theoretically reproduce the observed indicators of the chemical history of the solar neighborhood. In particular, the metallicity distribution of G-type dwarfs (i.e. the prevalence of such stars with respective iron content) is such an indicator, which is additionally critically affected by the type Ia supernova rate throughout the history of the Galaxy, and thus by the assumptions about their progenitors. This influence is greater than that of any known or conceivable uncertainty or oversimplification of our monolithic galactic chemical evolution model, which is used in combination with a population synthesis code including detailed binary evolution. We applied various assumptions and parameters concerning both binary and galactic evolution. Additionally, we updated and extended the study by \citet{mennekens2010} concerning the ability of both progenitor scenarios to reproduce the observed type Ia supernova delay time distribution (the evolution of the number of such events as a function of time after starburst) in passively evolving elliptical galaxies.

We conclude that both the delay time distribution and the G-dwarf metallicity distribution point towards a significant contribution by both the single degenerate and double degenerate channel. Under the assumption that their individual iron yields are of the same order of magnitude, the ability of either traditional progenitor scenario to be solely responsible for all type Ia supernova induced iron enrichment is ruled out. The best match with the observed G-dwarf metallicity distribution (and other galactic observables) is found when assuming that all double C-O white dwarf mergers explode as type Ia supernovae, as well as using a slightly larger convective core mass to account for rotational effects. In addition, the critical dependence of both distributions on certain binary and galactic evolutionary processes, exactly those processes which are modeled with still uncertain parameters in population synthesis studies, is a way to learn more about these processes and thus further constrain these parameters. In particular, we find that most double degenerate type Ia supernova progenitors need to go through a quasi-conservative, stable Roche lobe overflow phase, followed by a common envelope evolution. On a galactic level, a satisfying reproduction of the G-dwarf metallicity distribution requires a high binary frequency and a star formation rate that is relatively constant in time.

\section*{Acknowledgements}
      We thank Ivan Ram\'irez for a useful discussion regarding our comparison with observational AMRs, as well as Joke Claeys, Noam Soker and Ashley Ruiter for constructive comments on the manuscript.

\end{document}